\documentclass[aps,prl,twocolumn,floatfix]{revtex4}
\usepackage{bm}
\usepackage{epsf}
\usepackage{amssymb}
\usepackage{amsmath}
\usepackage{graphicx}
\usepackage{rotating}
\usepackage{epsfig}
\usepackage{psfrag}
\usepackage{amsmath}
\usepackage{hyperref}
\usepackage{subfigure}
\newcommand{\bk}{{\bf k}}

\newcommand{\pmat}[1]{\begin{pmatrix}
#1
\end{pmatrix}}
\newcommand{\dg}{^{\dagger}}

\newcommand{\up}{\uparrow}
\newcommand{\dw}{\downarrow}
\begin{document}

\title{Kondo Breakdown in Topological Kondo Insulators}

\author{Victor Alexandrov$^1$, Piers Coleman$^{1,2}$ and Onur Erten$^1$}
\affiliation{$^1$Center for Materials Theory, Rutgers University, Piscataway, New Jersey, 08854, USA \\ $^2$Department of Physics, Royal Holloway, University of London, Egham, Surrey TW20 0EX, UK }

\begin{abstract}
Motivated by the observation of 
light surface states in SmB$_6$, we examine the effects of surface Kondo
breakdown in topological Kondo insulators. We present both numerical
and analytic results which show that the decoupling of the localized
moments at the surface disturbs the compensation between light and 
heavy electrons and dopes the Dirac cone. 
Dispersion of these uncompensated surface states are dominated 
by inter-site hopping, which leads to a much lighter quasiparticles. These surface states 
are also highly durable against the effects of surface magnetism and decreasing
thickness of the sample. 
\end{abstract}
\maketitle

Kondo insulators 
are a class of strongly
correlated electron material in which the screening of local moments
by conduction electrons 
gives rise to 
an insulating gap at low temperatures
\cite{fiskaeppli,Coleman_book}.
The first Kondo insulator, 
SmB$_6$, discovered more than 40 years ago\cite{Menth_PRL1968}, 
has  attracted renewed interest 
due to its unusual surface transport properties:
while its insulating 
gap develops around $T_K\simeq
50$K, the resistivity saturates below a few Kelvin\cite{Allen_PRB1979}. 
Although this excess conductivity was 
originally ascribed to 
mid-gap impurity states\cite{Nickerson_PRB1971}, renewed interest
derives from the proposal that SmB$_{6}$ 
is a topological Kondo insulator, developing protected conducting
surface states at low temperatures
\cite{Dzero_PRL2010, Dzero_PRB2012,Alexandrov_PRL2013, Fu_PRL2013}. 
Experiments\cite{Wolgast_PRB2013,
Zhang_PRX2013, Kim_SciRep2013, Kim_NatMat2014} have since confirmed 
that the plateau conductivity derives from surface states, and these 
states  have 
been resolved by 
angle-resolved photoemission spectroscopy
(ARPES)
\cite{Jiang_NatComm2013,Neupane_NatComm2013, Xu_PRB2013,Frantzeskakis_PRB2013}.  
The most recent spin-ARPES experiments 
have also resolved the helicoidal 
spin polarization of the surface quasiparticles
expected from topologically protected 
Dirac cones\cite{Xi_NatComm2014}. 

Yet, despite this success, certain aspects of these materials remain
unexplained. A particularly notable problem, is that 
both quantum
oscillation\cite{Xiang_arxiv2013} and ARPES\cite{Neupane_NatComm2013,
Jiang_NatComm2013, Xu_PRB2013, Frantzeskakis_PRB2013, Kim_NatMat2014,
Xi_NatComm2014} studies show that the surface quasiparticles are
{\sl light}, with Fermi velocities  $v_s$, ranging from $220
{\rm meV \AA}$\cite{Jiang_NatComm2013} to $300 ~{\rm meV
\AA}$\cite{Neupane_NatComm2013}; yet current 
theories\cite{Dzero_PRL2010, Dzero_PRB2012,Alexandrov_PRL2013,
Fu_PRL2013} predict {\sl heavy} Dirac quasiparticles with 
velocities $v_s \sim 30-50~{\rm meV \AA}$, 
an order of magnitude slower.

\begin{figure}[t!]  \centerline{
\includegraphics[width=8.5cm]{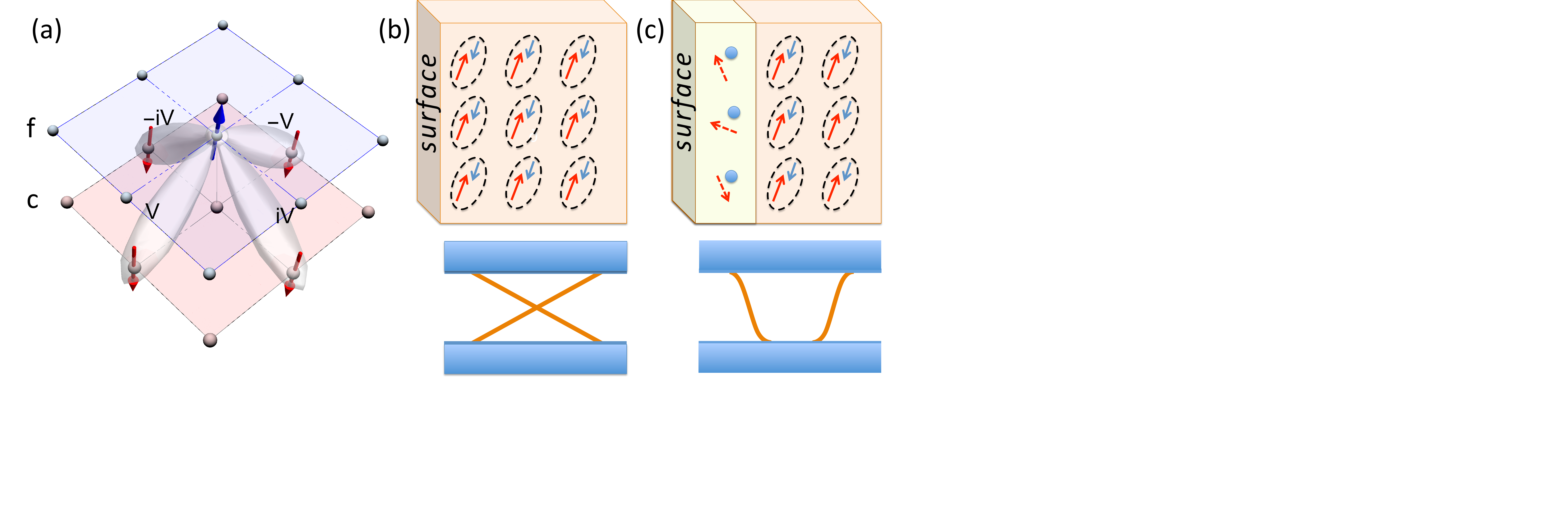} } \caption{(a)
Schematic showing the the non-local Kondo interaction at the surface of
topological Kondo insulator. 
The opposite parities 
of the $f$ and $d$ bands cause the hybridization to acquire
a p-wave, spin-dependent form
factor ${\Phi(\vec{R})}\propto i \vec{R}\cdot \vec{\sigma
}$ (see eq. \ref{phi_def}), which
vanishes onsite.
(b) and (c) respectively contrast the surface states
of a topological 
Kondo insulator, before and after surface Kondo breakdown, showing
how the Dirac point drops into the valence band, giving rise
to lighter, faster surface states.}  \label{Fig:1}
\end{figure}

Here we propose a resolution to
this problem, by taking into account of the breakdown of the 
Kondo effect at the surface. 
The essence of our theory is based on the observation
that the reduced co-ordination of the  Sm$^{3+}$ ions
at the surface 
causes a marked reduction in the surface Kondo temperature $T_{K}^{s}\sim T_{K}/10$,
so that the screening of local moments at the surface
is either suppressed to much lower temperatures, 
or fails completely due to an intervention of surface magnetic order. 
``Kondo breakdown''\cite{Si_Nat2001, Coleman_JPhys2001, Senthil_PRL2003, Pepin_PRL2007, Paul_PRL2007} liberates unquenched moments at the surface, and
has the effect of 
shifting  the mixed valence of the surface Sm ions towards
the higher entropy, $4f^{5}$ (3+) configuration of the unquenched
moments, as observed in X-ray absorption
spectroscopy \cite{Phelan_PRX2014}. 
Most importantly, Kondo breakdown
will liberate  a large number of carriers previously bound inside 
Kondo singlets at the surface (see Fig1(b) and (c)):
this process 
localizes $4f$-holes, driving up the electron count
in the surface states to form
large spin-polarized Fermi surfaces determined by the Luttinger sum rule
\begin{eqnarray}
\frac
{\mathcal{A}_{\hbox{\small FS}}}{(2\pi)^{2}} = \Delta n_f
\label{luttinger}
\end{eqnarray}
where $\mathcal{A}_{FS}$ is the total Fermi surface area 
of the singly-degenerate surface states, 
$\Delta n_f$ is the change the $f$ valence, {\sl i.e.} for Sm $4f^{5.6+}\rightarrow 4f^{6+}$,
$\Delta n_f = 0.4$, in units where the lattice constant $a=1$. 
A detailed analysis presented
later, shows
that the dispersion of these highly doped surface states 
 are dominated by a quadratic inter site hopping term 
\begin{equation}
E(k_x, k_y) = E_0 +\sqrt{T_k t_f}r(\delta)k_\perp+  t_f p(\delta)k_\perp^2,
\label{eq:quadratic}
\end{equation} 
where $t_f$ is the effective $f$ hopping, while $r(\delta)$ and
$p(\delta)$ are functions that depend on the surface scattering phase
shift $\delta$, as defined later.
The quadratic term, absent in a conventional topological surface
state, 
appears when the perfect compensation
between heavy and light electrons is  disrupted at the surface by Kondo
breakdown: this term increases the velocity of the
surface states by about an order of magnitude, providing a natural
explanation to the light surface states observed in
experiments. Moreover, the heavy doping of the surface
states shifts the Dirac point into the valence band, leading to an
unusual protection of surface states against decreasing thickness and
surface magnetism. We will show that the
interaction of the local moments and the light
surface states is described by a new kind of ``chiral Kondo
lattice'' with the potential for a rich phase diagram of competing 
interacting surface states. 

\begin{figure*}[t!]
\centerline{
\includegraphics[width=17cm]{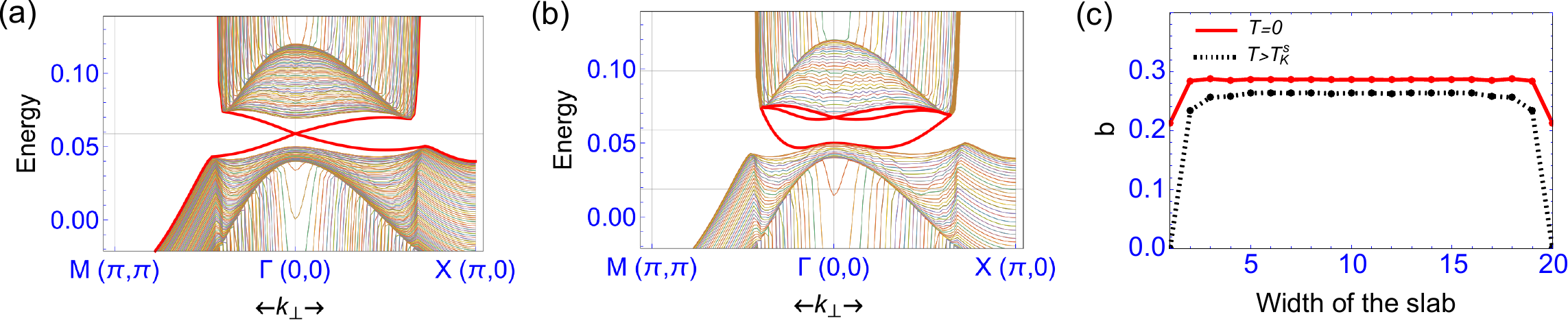}
}
\caption{Slab calculation for topological Kondo insulator: (a) for the uniform solution, (b) surface Kondo breakdown: surface local moments decouples from the rest of the system. Notice the large Fermi surface with high velocity quasiparticles in the case of the surface Kondo breakdown. (c) Inhomogeneous mean field theory calculations for the slave boson amplitude, $b$ as a function of the thickness of the slab. $b$ is reduced at the surface and for $T>T_K^s$, the local moments completely decouple.}
\label{Fig:2}
\end{figure*}

\noindent {\sl Model:} We
introduce a simplified lattice model for topological Kondo insulators,
described by 
the periodic
Anderson model:
\begin{widetext}
\begin{eqnarray}\label{l}
H& =& \sum_{(i,j) \alpha ,\beta } \left(d\dg_{i \beta}, f\dg_{i
\beta } \right)
\pmat{
-t_{ij}- \mu\delta_{ij}&
\tilde{V} \Phi_{\beta \gamma} (\vec{R}_{i}-\vec{R}_{j})
\cr
\tilde{V} \Phi_{\beta \gamma} (\vec{R}_{i}-\vec{R}_{j})
& \tilde{t}^f_{ij} + \tilde{\epsilon}_{f}\delta_{ij}}
\pmat{d_{j \gamma}\cr f_{j\gamma }}
 + U\sum_i n_{if\uparrow}n_{if\downarrow}
\end{eqnarray}
\end{widetext}
where the $d^\dagger_{j\alpha }$
and $f^\dagger_{j\alpha }$ respectively create 
conduction and f electrons at site $j$, with corresponding 
hopping matrix elements $-t_{ij}$ and $\tilde{t}_{ij}^{f}$.
$U$ is the onsite Coulomb repulsion of f-electrons. 
$\tilde{V}$ and $\tilde{\epsilon}_{f}$ are the
bare hybridization and $f$ level position. The form-factor
\begin{equation}\label{phi_def}
\Phi (\vec{R} )_{\beta \gamma}=  \left\{\begin{array}{cc}
-i \hat  R\cdot 
\left(\frac{\vec{\sigma }}{2}\right)_{\beta \gamma }
, & \hbox{n.n.}\cr
0 & \hbox{otherwise}
\end{array}
\right.
\end{equation}
describes the spin-orbit coupled hybridization between neighboring $f$ and 
$d$ electrons. 
This simplified form factor 
captures the important 
vectorial spatial structure of hybridization between f and d
states, whose orbital angular momenta differ by one unit.
The hybridization is odd-parity and vanishes onsite
because the heavy and light orbitals have opposite spatial parity. 
Fig. 1(a) shows a schematic of our model. Note that it 
explicitly includes an $f$ dispersion\cite{Coleman_book}, which is
required for a fully gapped spectrum. 
Microscopically these terms are expected to 
result from the indirect hopping of f-electrons via the p-orbitals
of the Boron ions. 

To describe the large $U$ limit 
we adopt a slave-boson mean-field theory
and carry out a saddle point approximation in the bulk, 
giving rise to the mean-field
Hamiltonian \cite{Coleman_PRB1987}.
\begin{eqnarray}\label{MF_Hamiltonian}
H_{MF}& =& \sum_{\bk \beta,\gamma } \left(d\dg_{\bk  \beta }, f\dg_{\bk
\beta } \right)
{
\pmat{
\epsilon_{\bk } -\mu&
V \vec{s}_{\bk }\cdot\vec{\sigma}_{\beta \gamma}\cr
V\vec{s}_{\bk }\cdot \vec{\sigma}_{\beta \gamma}
& \epsilon_{f\bk }+\lambda}}
\pmat{d_{\bk \gamma }\cr f_{\bk
\gamma}}\cr
&
 +& {\cal N}_{s} (\lambda-\tilde{\epsilon}_{f})\left(|b|^{2}-Q  \right).
\end{eqnarray}
Here,
$V= \tilde{V}b$ is the renormalized hybridization, 
where $b$ is the slave boson projection amplitude. 
The f-hopping now becomes $t^{f} = b^2 \tilde{t}^f$. We consider a general
nearest neighbor and next nearest neighbor dispersion, 
$\epsilon_k = -2t\sum_i\cos k_i-4t^\prime \sum_{i\neq j} \cos k_i \cos k_j $ while
$\epsilon_{f\bk } = -\alpha \epsilon_k$, where, for simplicity, 
we have taken the ratio between the 
$f-$- and $d-$ electron hoppings  to be a single  fixed constant
$\alpha = t_f/t= t_{f}^{\prime}/t^{\prime}$ 
for both nearest neighbor and second-neartest neighbor
hoppings. 
The quantity $\lambda$ is the constraint field 
that imposes the mean-field constraint $Q=n_f+b^{2}$, 
where $Q$ is the local conserved charge associated with the slave boson
treatment of the infinite $U$ limit, here taken to be $Q=1$;
${\cal N}_{s}$ is the total number of sites. 
In momentum space the 
form factor of the hybridization now takes 
the form $\Phi (\bk )=\vec{s}_{\bk }\cdot \vec{\sigma}$, where $\vec{s}_{\bk } =
(\sin{k_x}, \sin{k_y}, \sin{k_z})$, which reduces to $ \bk\cdot \sigma$
at small k.  This odd-parity hybridization is reminiscent of the gap
function in topological superfluid $^{3}$He-3B.
This model can be regarded as an adiabatic continuation 
from small to large $U$ at infinite spin-orbit coupling.
Band crossing between the odd and even parity bands generates the 
topological band structure\cite{Hasan_RMP2010, Qi_RMP2011} with 
protected  surface states.

In order to obtain the surface state spectrum, we adopt Volovik's 
approach, mapping the reflection at the boundary 
onto transmission through an interface where the hybridization changes
sign\cite{Volovik2009}. This method enables us to solve for the surface states
using a linearized Hamiltonian. 
For pedagogical purposes, we take $t^\prime = t_f^\prime = 0$. 
We can treat 
the surface eigenstate at $k_{x,y}=0$ as a one-dimensional problem, and
since the main surface scattering 
effect takes place close to the
Fermi surface,
we can linearize the 
dispersion(\ref{MF_Hamiltonian})  normal to the surface and obtain
the surface eigenstates at $k_{x,y}=0$.
The two interface
eigenstates at $k_{x,y}=0$
are given by
\begin{equation}\label{wavefunction}
\psi_\pm \sim \left[ \begin{array}{c}
1\pm 1\\1\mp 1
\end{array}\right]_\sigma \left[ \begin{array}{c}
\sqrt{\alpha}\\\pm i
\end{array}\right]_\tau e^{-i k_z z -\kappa |z|}
\end{equation}
where $k_z = \lambda/ [t(\alpha+1)]$ and $\kappa =
V/(\sqrt{\alpha}t)$. $\tau$ and $\sigma$ denote the orbital and the spin component of the wave function. The corresponding energy of a state at the Dirac point is
$E_0 =\lambda/(\alpha +1) $. 
The transverse dispersion that develops at finite $k_{x,y}$
is then treated using first order degenerate perturbation theory, 
projecting the full Hamiltonian $H_{MF}$ onto
the surface bound-states, $H_{eff} =\langle
\psi_\pm|H_{MF}|\psi_\pm\rangle$. This leads to the dispersion 
\begin{equation}\label{Dispersion1}
E(k_x,k_y) = {\lambda\over \alpha+1}+ 2\frac{V\sqrt{\alpha}}{\alpha +1 }k_\perp+ \mathcal{O}(k_\perp^3)
\end{equation}
where $k_\perp= \sqrt{k_x^2+k_y^2}$. Since $\alpha \ll 1$, the surface
states are composed of heavy quasiparticles. Note that at this stage, 
the quadratic $k^2$ term is absent. 

\noindent {\it Surface Kondo breakdown:} 
The reduced
co-ordination number of the f-electrons, plus the internal gapping of
the bulk states  lowers the effective Kondo coupling constant. 
If we assume crudely that the Kondo coupling constant is reduced by a
factor of $5/6$th, the reduced Kondo temperature is then 
\begin{equation}\label{surface_Kondo}
T_{K}= D e^{- \frac{1}{J\rho }}\rightarrow T_{K}^{s}=D \exp (- \frac{6}{5J\rho
}) = T_{K }\left(\frac{T_{K}}{D} \right)^{1/5}
\end{equation}
For example, if we take the bulk Kondo temperature
$T_{K}\sim 50K$ with half band-width $D\sim 2eV\sim
2\times 10^{4}K$, this gives a surface Kondo temperature of 
$T_{K}^{s}\sim 15K$. In practice, the
gapping of the bulk states makes this an upper bound. 
This effect has been recently explored in
1D\cite{Alexandrov_PRB2014, Lobos_arxiv2014}. However, unlike one dimension, where
the spin-decoupled phase
is no longer topological\cite{Alexandrov_PRB2014},  in two and three
dimensions, the topological insulator is protected by time reversal
symmetry and the spin-decoupled state remains topological in the
bulk. With these considerations in mind, we model the
surface Kondo breakdown as the suppression of the slave boson
amplitude $b$ to zero on the surface layer of the TKI. 

Surface Kondo breakdown localizes mobile $f$-quasiparticles at the
surface, removing them from the Fermi sea.
In SmB$_{6}$, this is equivalent 
to a shift in the effective f-valence from 
$4f^{5.6} \rightarrow 4f^{6}$ at the surface. This is a reduction in
the number of f-holes, that corresponds to an increase $\Delta n_f=0.4$
in the density of electrons on the surface. From Luttinger's
sum rule (eq.~\ref{luttinger}), the area enclosed by the Fermi surface
is equal to the total electron density. The Fermi surface area observed in
ARPES is 0.3-0.35\cite{Neupane_NatComm2013, Jiang_NatComm2013,
Xu_PRB2013, Frantzeskakis_PRB2013, Kim_NatMat2014, Xi_NatComm2014}, in
agreement with Luttinger sum rule. 

Next we examine the effect of Kondo breakdown on the
dispersion of the surface states. 
The effect of Kondo breakdown can be incorporated by
introducing a  scattering phase shift $\delta = k_za$ 
into the $f$ component of the surface state wave function, 
which ensures that surface node of the 
$f$ part of the wave function is shifted one lattice unit in towards
the bulk, depleting the surface of mobile f-electrons (see 
supplementary information for details\cite{supp}).
In Volovik's approach, $\delta $ becomes a transmission phase shift. 
The unperturbed energy of the surface bound-state 
is now modified $E_0^\prime = (\lambda+ 2
V\sqrt{\alpha} \sin \delta)/(1+\alpha) $ and the dispersion becomes
\begin{equation}\label{Dispersion2}
E^\prime(k_x, k_y) = E_0^\prime +V\sqrt{\alpha} r(\delta)k_\perp+   \alpha t\frac{e^{\frac{2 \kappa \delta}{k_z}}-1}{\alpha e^{\frac{2 \kappa \delta}{k_z}}+1} k_\perp^2
\end{equation}
Note the appearance of a $k_\perp^2$ term, driven 
by an increase the conduction electron content of the
surface states. The subleading linear term is also modified, containing
the renormalization $r(\delta)$ of order one: $r(\delta) =2 (\cos \delta
+\frac{\kappa}{k_z} \sin \delta)/(1+\alpha \exp(\frac{2 \kappa
\delta}{k_z}))$. Substituting $\alpha t = t_f$ and $V\sim \sqrt{T_K t}$ in
eq.~\ref{Dispersion2} leads to eq.~\ref{eq:quadratic}.  This 
analytic  result provides a natural explanation
to the light quasiparticles in SmB$_6$. 
For a parameter choice of
$t=500~{\rm meV}$, $t_f=5~{\rm meV}$, consistent with band structure
calculations\cite{Fu_PRL2013}, and the value of $V=100~{\rm meV}$
which leads to the experimentally observed gap $\Delta \simeq 20~ {\rm
meV}$ we obtain surface state velocity $v_{ss}\sim 58 ~{\rm meV \AA}$
without surface Kondo breakdown and $v_{ss}\sim 340 ~{\rm meV \AA}$
with surface Kondo breakdown in agreement with the $v_{ss}\sim300~{\rm
meV \AA}$ measured by ARPES
experiments\cite{Neupane_NatComm2013}. Here the quoted surface velocities 
were 
calculated at the mid-gap  energy
of the Dirac point in the absence of Kondo breakdown. 
We can not however, account for
the much larger 
surface velocity 
$v_{ss}\sim 4000-7000$meV$\AA$ reported in quantum oscillations
\cite{Xiang_arxiv2013}. 

To confirm the results of our simplified analytical calculations, we
have carried out a series of slab calculations with and without Kondo
breakdown. 
As a first check, we simulated Kondo breakdown by setting
the amplitude of the surface slave boson field to zero. 
Without surface Kondo
breakdown (Fig 2(a)), the surface states are heavy, but as expected, 
when the slave boson amplitude is suppressed to zero on the surface, 
the topological states become much lighter as shown in 
Fig. 2(b). Moreover, the area of the light Fermi
surface is significantly enhanced, corresponding to the extra density of
carriers on the surface.

Next, to confirm that the assumption of surface Kondo breakdown 
we performed a fully self-consistent
mean field calculation in which both $b (z)$ and
$\lambda (z)$ were allowed to 
vary along the $z$ direction.  
In this self-consistent calculation, we chose parameters where 
the heavy and light bands cross at the
three X-points, as seen SmB$_{6}$
\cite{Jiang_NatComm2013,Neupane_NatComm2013, Xu_PRB2013,Frantzeskakis_PRB2013};
from a technical stand-point, the crossing at the three $X$ points
is required to enhance the f-character, achieving the Kondo limit
$n_{f}\sim 1$ while
still in the strong topological insulator phase. We achieve this
crossing in our model by taking $t=-t^\prime/2$ , such that the heavy
and light bands cross at the 3 different $X$ points.  
We note that for the simpler
choice $t^\prime = 0$, bands cross at the $\Gamma$ point, restricting
the strong topological insulating behavior to the region to $n_f \ll 1$.  
Fig 2 (c) presents the results of these calculations. We find 
as expected, that the self-consistently determined 
slave boson amplitude $b$ is depressed
at the surface, corresponding to a reduction of the surface Kondo
temperature. In our model calculation, for which $T_{K}/D\sim 0.025$
found that $b^{2}$ is depressed by $1/2$ at the surface, corresponding to
$T_{K}^{s}/T_{K}\approx 0.5 \sim (T_{K}/D)^{1/5}$, a result consistent
with equation (8).  We also found that 
as the temperature is raised above the 
the surface Kondo temperature $T_K^s$, 
the surface $b$ collapses to zero, even though it is still finite in
the bulk. The 
dispersion 
displays a corresponding transition
between heavy and light surface states as the temperature
is raised through $T_{K}^{s}$.  In practice, the possible intervention of magnetism 
or other instabilities may permanently prevent the low-temperature 
re-establishment of the Kondo effect at the surface. 

\noindent{\it Chiral Kondo lattice:} The consequences of surface
Kondo breakdown are rather interesting. The decoupled f-electrons
now interact with the chiral surface states to form 
a new kind of Kondo lattice
in which the conduction sea of chiral electrons
is now {\sl singly} degenerate, 
interacting with the local moments via a
Hamiltonian  
\begin{eqnarray}
H_{CKL} = \sum_{k} \epsilon_c(k) c_k^\dagger c_k \nonumber  + J
\sum_{i}\psi \dg_{i}\vec{ \sigma }\psi_{i} \cdot \vec{S}_{i}+ J_H \sum_{\langle ij \rangle}\vec{S}_i \cdot \vec{S}_j
\end{eqnarray}
where $J$ is the Kondo coupling, 
${\vec S}_j$ is the decoupled local moment and
where $c^\dagger_{k}$ creates a chiral, spin polarized surface states whose dispersion 
$\epsilon_c(k)$ is given in eq.~\ref{Dispersion2}.
The localized two-component  electron field at site $j$, 
\begin{equation}\label{}
\psi_{j} = \frac{1}{\sqrt{2{\cal  N}_{s}}}\sum_{\bk }\begin{pmatrix}
1\cr \hat k_{x}+i \hat k_{y} 
\end{pmatrix}
c_{\bk }e^{- i \bk  \cdot {\bf R}_{j}}
\end{equation}
is constructed from the spin-polarized surface states by combining
orbital and spin angular momentum. $J_H \sim
4t_f^2/\tilde{\epsilon}_f$ is the antiferromagnetic Heisenberg
exchange between the local moments which is derived by a
Schrieffer-Wolff transformation of eq. \ref{l}. The ground-state of
the chiral Kondo lattice can be either magnetically ordered state or a
heavy fermion liquid where the local moments are screened by chiral
fermions.  As in conventional heavy fermion systems, near the quantum
critical point that separates these two limits, there is the
possibility of strange metal behavior and superconducting
ground-states. A 2D version of a similar model with 1D edge states
have been explored recently\cite{Altshuler_PRL2013}.

Even the magnetic phase diagram of this lattice is expected to be
rich. The  RKKY interaction in 
a chiral Kondo lattice gives rise to Heisenberg,
Dzyaloshniskii-Moriya and compass anisotropy terms
\cite{Zhu_PRL2011} of strength $J$, $D$ and $A$ respectively. 
For $AJ/D^2>1$ the ground state is an in-plane
ferromagnet whereas for $AJ/D^2<1$, it is a spiral of variable pitch
set by $D/J$\cite{Banerjee_NatPhys2013}. Such a spiral
state evolves into a skyrmion crystal in applied magnetic
field\cite{Banerjee_PRX2014}. Experimentally, in SmB$_{6}$, 
indications of magnetism have been
observed as hysteresis in magnetoresistance
experiments\cite{Nakajima_arxiv2013, Eo_arxiv2014}. The absence of ordered
ferromagnetic moment in XMCD experiments\cite{Phelan_PRX2014} suggests
that the ground state might be a spiral with no net moment. Magnetic
order breaks the time reversal symmetry and can in
principle gap the surface states. 
We note that of the most recent
single crystals show no plateau in
resistivity\cite{Phelan_PRX2014} indicating that magnetism may 
play a crucial role in the topological behavior.  
Better sample quality and tuning the chiral Kondo lattice
through a quantum critical point opens up the possibility of exotic
superconductivity on the surface of topological Kondo insulators.

\noindent {\it Thickness dependence of surface states:} Perfect
topological protection of the surface states requires an infinite
bulk. When the thickness of the system is comparable to the decay
length, $\xi = \kappa^{-1}$ of the surface states, the two surface
states on each surface mix, giving rise to a gap in the
spectrum. Since the mixing is a zero-momentum transfer process, the
gap opens at the doubly degenerate Dirac point as shown in
Fig.~3(a). This effect has been observed in conventional topological
insulators like $Bi_2Se_3$ where the surface states get gapped for
thickness 3-5 quintuple layers\cite{Zhang_NatPhys2010}. However with
surface Kondo breakdown, the Dirac point shifts in the valence band,
giving rise to a unique protection against decreasing thickness of the
sample (Fig.~3(b)). Recent transport experiments on films about
L=100-200nm thicknesses observe topological surface states
\cite{Yong_arxiv2014}. We predict that the topological states will
persist even in ultra-thin samples, $L \sim \xi \sim 10nm$, making
$SmB_6$ a perfect candidate for thin film applications.  This
mechanism also gives rises to protection against small time reversal
symmetry breaking effects such as weak magnetism at the surface.

\begin{figure}[t!]
\centerline{
\includegraphics[width=7cm]{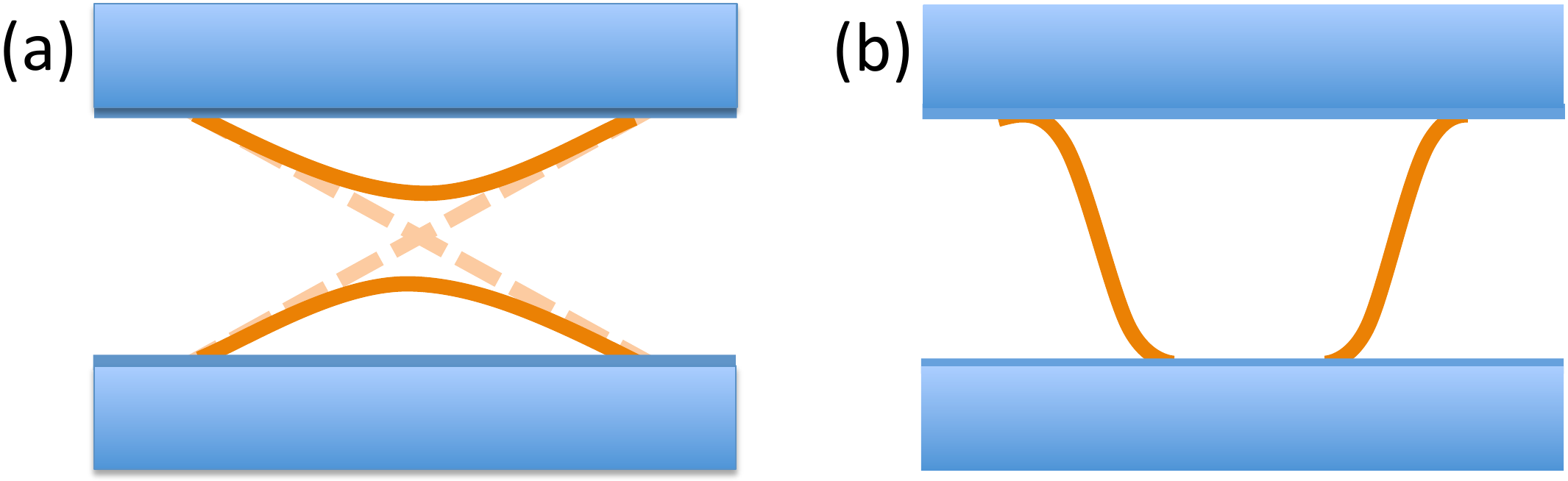}
}
\caption{(a) Without the surface Kondo breakdown, the surface states are susceptible to reducing thickness and and effects of magnetism where an insulating gap starts developing around the Dirac point. (b) In the case of surface Kondo breakdown, the Dirac point sinks in the valence band, providing an unusual protection to the surface states. The spectrum stays metallic and qualitatively unchanged for moderate values of perturbation.}
\label{Fig:3}
\end{figure}

\noindent {\sl Conclusion:} To conclude, we have addressed the origin
of the light surface states of topological
Kondo insulators as a simple consequence of surface Kondo
breakdown. Our results appear to reconcile the observation of light
surface states and the peculiar shift of valence in $Sm$ observed in
X-ray measurements. This theory  also predicts an 
increased stability of the the surface states against
finite thickness and magnetism may be important in thin film
device fabrication. We have also argued that the interaction of the
surface spins with spin-polarized 
light surface states is described by a chiral Kondo lattice: a two
dimensional Kondo lattice with strong parity violation, with the
potential for a rich phase diagram in the vicinity of its magnetic
quantum critical point. 
The detailed character of the surface states at low temperatures, with
the possibility of a wide variety of magnetic or other orderings
\cite{Efimkin_PRB2014, Roy_arxiv2014} remains a fascinating topic for future research.

\noindent
{\it Note added:} We recently became aware of low temperature magneto thermo-electric transport measurements\cite{Luo_PRB2015} 
which indicate the development of a very heavy surface Fermi liquid below 3K, with quasiparticle group velocities in 
the range $5-10 {\rm meV \AA}$, consistent with the development of a low-temperature surface Kondo effect.

\noindent
{\it Acknowledgments:} We gratefully acknowledge stimulating
conversations with Jim Allen, Tzen Ong, Kai Sun, George Sawatzky and David Vanderbilt. 
This work is  supported by Department of Energy grant DE-FG02-99ER45790.

\bibliographystyle{apsrev}

\pagebreak
\widetext
\begin{center}
\textbf{\large Supplementary material for `Kondo Breakdown In Topological Kondo Insulators'}
\end{center}
\setcounter{equation}{0}
\setcounter{figure}{0}
\setcounter{table}{0}
\setcounter{page}{1}
\makeatletter
\renewcommand{\theequation}{S\arabic{equation}}
\renewcommand{\thefigure}{S\arabic{figure}}
\renewcommand{\bibnumfmt}[1]{[S#1]}
\renewcommand{\citenumfont}[1]{S#1}

This supplementary material presents 
the details of the surface state calculations. We first
describe the Volovik method\cite{Volovik2009} and then apply it to our
model, with and without Kondo breakdown.
\section{Volovik method} 

Surface states are evanescent solutions of the
Hamiltonian which are nucleated by the boundary. The 
surface state wave function in the presence of a hard wall
consists of an incoming and an outgoing wave moving in opposite directions
whose superposition
vanishes at the boundary as shown in Fig 1(a). Volovik's
method\cite{Volovik2009} ``unfolds'' the reflection process into a 
transmission problem, with incoming and outgoing waves moving in the
same direction (see Fig. 1b).
Similar ``unfolding'' techniques have been
applied in wide variety of problems including the Kondo impurity
model\cite{revbethekondo}.
The Volovik transformation is made by applying the
reflection operator to the eigenvalue equation governing the 
incoming waves: under this 
a transformation, the odd parity
terms in the Hamiltonian reverse sign at the boundary:
in our particular, model the only odd parity term in the Hamiltonian is the hybridization, $V(z)$,
which in the Volovik mapping now 
changes sign at the boundary $V(z) = V {\rm sgn}(z) $ as
shown in Fig1(b). 
The Volovik method has the advantage that the ingoing and outgoing
waves carry the same momentum; it also 
brings out the close
analogy between surface states and Jackiw-Rebbi domain wall
bound-states, with 
the additional technical advantage that it can handle cases where $k_{z}\rightarrow 0$.

\begin{figure}[h!]  \centerline{
\includegraphics[width=12cm]{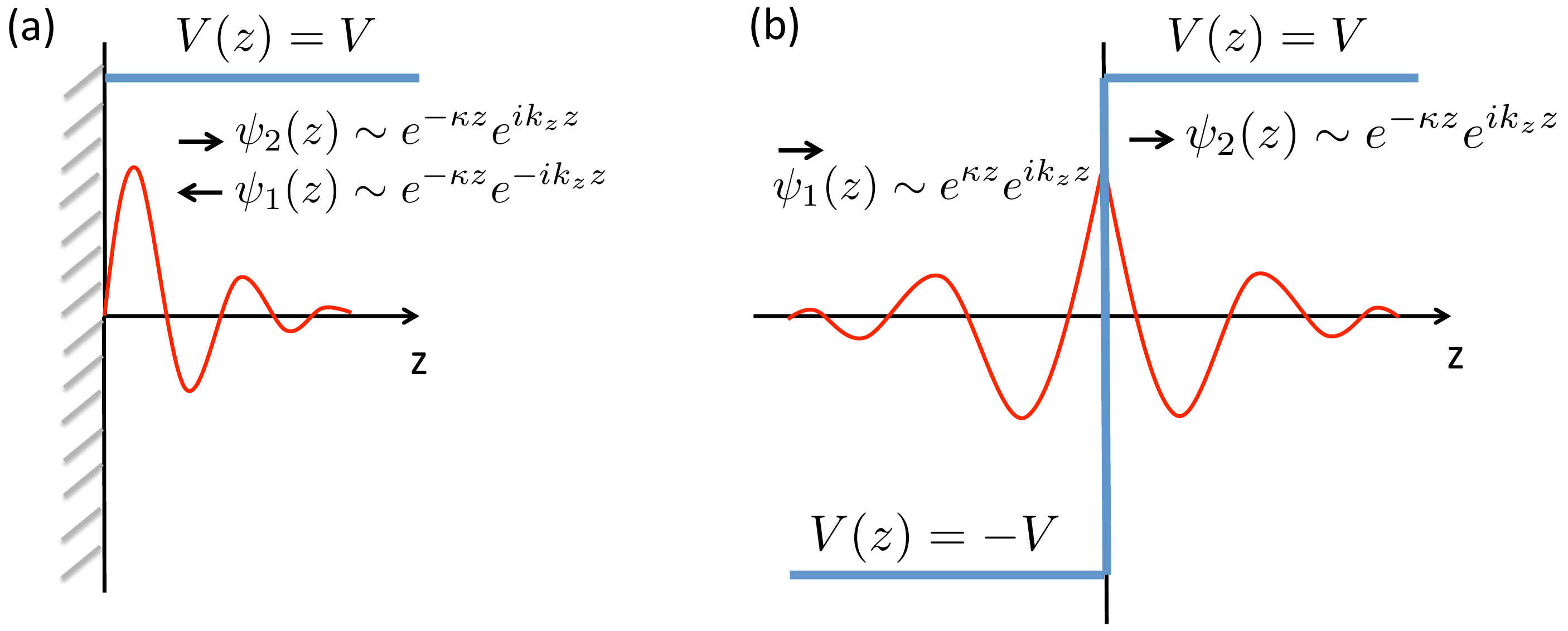} } \caption{Schematic of the bound state calculation (a) A hard wall with incoming and outgoing waves can be mapped onto (b) a reflection problem with a sign change in the hybridization.}  \label{Fig:1}
\end{figure}

To be more specific, consider first a reflection boundary problem. The
wavefunction consists of incoming and outgoing waves
\begin{equation}\label{}
\hbox{Reflection}\left\{\begin{array}{rcl}
\psi_{1,2} (z)&\sim &e^{-\kappa_{z}z}e^{\mp ik_{z}z}\chi\cr
H_{+}\psi_{1,2}&=& E\psi_{1,2} .
\end{array}\right.\end{equation}
Here $\psi_{1,2} (z)$ are eigenstates of the bulk Hamiltonian $H_{+}$
defined on the half-line $z>0$ and $\chi$ is a spinor. The
full wavefunction is given by
\begin{equation}\label{}
\psi  (z) = \psi_{1} (z)- \psi_{2} (z), \qquad \qquad  (z>0). 
\end{equation}
The boundary condition $\psi (z=0)=0$ is satisfied by the fact that
$\psi_{1} (0)= \psi_{2} (0)$. 

The Volovik method obtains an equivalent 
transmission problem by applying the parity
operator to the eigenvalue equation of the incoming wave
$PH_{+} \psi_{1} (z) = P E \psi_{1} (z)$, to obtain
\begin{equation}\label{}
H_{-}\psi_{1} (-z) = E \psi_{1} (-z)
\end{equation}
where $H_{-}= P H_{+}P\dg $ is the parity reversed Hamiltonian. The
corresponding transmission problem  is then given by 
\begin{equation}\label{}
\hbox{Transmission}\left\{\begin{array}{rcl}
\psi &=& \psi_{1} (-z)\theta (-z)+ \psi_{2} (z)\theta (z)\cr
H &=& H_{-}\theta (-z) + H_{+}\theta (z)
\end{array}
 \right.
\end{equation}
Here 
\begin{equation}\label{}
\psi (z)\sim e^{ik_{z}z}e^{-\kappa_{z}|z|}\chi 
\end{equation}
is a symmetric domain-wall bound-state. In our particular problem the
right and left hand Hamilonians are simply obtained by reversing the
sign of the hybridization:
$H_{\pm }= H (\pm V)$.

\section{Continuum model}
Our starting point is $H_{MF}$, the mean field Hamiltonian, eq. 5 in
the main text. We replace $k_z = (k_F + i \partial_z)$ and expand 
$H_{MF}$ to  quadratic order in $k$  so that
\begin{eqnarray}
H_{MF}= t(k^2 - k_F^2)\pmat{1& 0
\cr
0 & -\alpha}_{\tau}+V (z) (k_x \sigma_x +k_y \sigma_y+ k_z \sigma_z)\tau_1 +\lambda \pmat{0& 0
\cr
0 & 1}_{\tau}+ \mathcal{O}(k^3),
\end{eqnarray}
where
\begin{equation}\label{}
V (z)= V {\rm sgn} (z).
\end{equation}
Here the $\tau$- index refers to the conduction/f-electron 
degree of freedom, so that 
$\tau_{z}=+1$ describes a conduction electron while $\tau_{z}=-1$
describes an f-electron. The Pauli matrices $\sigma_{x,y,z}$ 
carry the spin degree of freedom.  We seek bound-state solutions
of the form
\begin{equation}\label{}
\psi (\vec{x}) = e^{i \vec{k}\cdot \vec{ x} - k |z|}\chi .
\end{equation}
where  $\vec{k}= (k_{x},k_{y},k_{z})$. 

We will be interested in the
small values of the transverse momenta, where $|k_{x,y}|\ll |k_{z}|$.
This allows us to expand the Hamiltonian in 
in powers of $k_{\perp }$ 
$H_{MF}=H_0+H_1+H_2+ \mathcal{O}(k^3)$, where
\begin{eqnarray}
&& H_0 =2t k_F\pmat{1& 0
\cr
0 & -\alpha}_{\tau} \partial_z+ V (z)  k_F\sigma_z\tau_1 +\lambda \pmat{0& 0
\cr
0 & 1}_{\tau}+ \mathcal{O}(k^3),
\\
&& H_1=V (z) (k_x \sigma_x +k_y \sigma_y)\tau_1,  
\\
&& H_2=t(k_x^2 +k_y^2) \pmat{1& 0
\cr
0 & -\alpha}_{\tau} + \mathcal{O}(k^3).
\end{eqnarray}
Our procedure is to first obtain the eigenvalues of the zeroth order
Hamiltonian, 
$H_0\psi=E\psi$ and then to carry out perturbation theory on the
remaining components $H_{1}+H_{2}$.  This corresponds to first solving
for the bound-state at  $\vec{k}_{\perp }=0$.

We need to treat positive and negative $z$  separately. 
In the positive half-plane, ($z>0$) the 
eigenvalue condition  $H_{0}\psi  = E\psi $ gives 
\begin{eqnarray}
E\psi =2t k_F\pmat{1& 0
\cr
0 & -\alpha}_{\tau} \partial_z \psi + V  k_F\sigma_z\tau_1\psi  +\lambda \pmat{0& 0
\cr 0 & 1}_{\tau }\psi 
\end{eqnarray}
We may simultaneously diagonalize both energy and $\sigma_{z}$,
defining ``up'' and ``down'' eigenstates
$\sigma_z \psi_{\uparrow,\downarrow } = \pm \psi$. 
For $\sigma_z = +1$, the reduced eigenvalue equation is 
\begin{equation}
\partial_z\psi_{\uparrow} =
- 
\frac{i}{2t \alpha k_F} \pmat{E \alpha& V \alpha k_F
\cr
-V k_F& - (E-\lambda)}  \psi_{\uparrow}
\end{equation}
where we have dropped the orbital index $\tau$ for clarity. The
solution for $z>0$
can be written in the form of an evanescent wave.
\begin{equation}
\psi_{\uparrow} (z>0) =\frac{1}{\mathcal{N} V k_F} \pmat{1 
\cr
0}_\sigma
\pmat{i Vk_F \alpha
\cr
i r -\sqrt{\alpha V^2k_F^2 - r^2}}_\tau \cdot e^{(i k_z - \kappa) z}   
\end{equation}
where $\mathcal{N}$ is a normalization constant and $k_z,\kappa$ and $r$ are defined as follows
\begin{eqnarray}
&&k_z =\frac{ r -\alpha E}{2t \alpha k_F},
\\
&& 2r = E(1+\alpha)-\lambda,
\\
&& \kappa = \frac{\sqrt{\alpha V^2 k_F^2 - r^2}}{2t \alpha k_F}. 
\end{eqnarray}
Similarly, the $\sigma_z = -1$ solution 
is 
\begin{equation}
\psi_{\downarrow} (z>0) =\frac{1}{N V k_F} \pmat{0 
	\cr
	1}_\sigma
\pmat{- i Vk_F \alpha
	\cr
	i r -\sqrt{\alpha V^2k_F^2 - r^2}}_\tau \cdot e^{(i k_z -\kappa) z}   
\end{equation}

Following the same procedure, the decaying solutions for $z<0$ are
\begin{eqnarray}\label{l}
\psi_{\uparrow}(z<0) &=&\frac{1}{N V k_F} \pmat{1 
	\cr
	0}_\sigma
\pmat{- i Vk_F \alpha
	\cr
	i r + \sqrt{\alpha V^2k_F^2 - r^2}}_\tau \cdot e^{(i k_z
	+\kappa) z}   \cr
\psi_{\downarrow}(z<0) &=&\frac{1}{N V k_F} \pmat{0 
	\cr
	1}_\sigma
\pmat{ i Vk_F \alpha
	\cr
	i r + \sqrt{\alpha V^2k_F^2 - r^2}}_\tau \cdot e^{(i k_z +\kappa) z}   .
\end{eqnarray}

\vspace{.4cm}
\subsection{Without Kondo breakdown}
We first consider the case the boundary conditions in the absence of Kondo breakdown. We need to match $\psi_{\uparrow}(z<0)$ and $\psi_{\uparrow}(z>0)$ at $z=0$. This gives the condition: $r = 0$. We then find $E_0$, $\kappa$ and $k_z$ for $r=0$.
\begin{eqnarray}\label{phizero}
&& E_0 = \frac{\lambda}{1+\alpha},
\\
&&  \kappa =\frac{V}{2t \sqrt{\alpha}},
\\
&& k_z =\frac{-\lambda \alpha}{2t (1+\alpha) \alpha k_F}.
\end{eqnarray}
So far we have considered the $\mu=0$ case. 
The solution for $\mu \neq 0$ can be obtained by
\begin{equation}
H(\lambda,\mu) = H(\lambda-\mu,0)+\mu.
\end{equation}

The remainder of the Hamiltonian is treated within perturbation
$\delta H = H_1+H_2$ for small $k_\perp$.  
The effective Hamiltonian derived by projecting
$\delta H$ on to the surface states: $H_{\sigma\sigma'} = 
\langle\psi_{\sigma}|\delta
H|\psi_{\sigma' }\rangle$, where $\sigma, \sigma '\in
\{\uparrow,\downarrow  \}$ are spin indices.
The quadratic part of the Hamiltonian $\delta H$ cancels
exactly: $\langle\psi_{\sigma }| H_2|\psi_{\sigma '}\rangle \equiv0$. 
The remaining linear
part gives
\begin{equation}
H_{\uparrow\downarrow } = \langle\psi_{\uparrow}| H_1|\psi_{\downarrow}\rangle = -2 i \sqrt{\alpha} \frac{V \alpha}{ \mathcal{N}^2} (k_x+i k_y),
\end{equation}
where the normalization $\mathcal{N}^2 =\int dz
\langle\psi_{\uparrow} |\psi_{\uparrow}\rangle$ 
takes the form $\mathcal{N}^2= \alpha(\alpha +1)$. The effective Hamiltonian is then
\begin{equation}
H_{eff} =H_0+\pmat{0& i k_+\cr -i k_-&0} \frac{ 2 \sqrt{\alpha} V}{\alpha +1}
\end{equation}
where $k_\pm = k_x \pm i k_y$ and the corresponding dispersion is
\begin{equation}
E_{eff} = E_0+ \frac{ 2 \sqrt{\alpha} V}{\alpha +1}k_\perp
\end{equation}
where $k_\perp$ is defined as $k_\perp^2 = k_x^2+k_y^2$.

\subsection{Kondo breakdown}
In the presence of Kondo breakdown, the hybridization vanishes at the first site. As a result, $c$ and $f$ bands are subject to different boundary conditions. This effect can be captured by a simple phase shift $\delta$, such that the boundary condition  $\psi_{\uparrow}(z>0)\big|_{z=0} = \psi_{\uparrow}(z<0)\big|_{z=0}$ is modified to
\begin{equation}
\psi_{\uparrow}(z>0)\big|_{z=0} = \pmat{1 &0
\cr
0&e^{2i\delta}}_\tau \psi_{\uparrow}(z<0)\big|_{z=0}
\end{equation}
Now the previous condition of r=0 is modified to $ r=\sqrt{\alpha}Vk_F \cos \delta$. Hence the energy and the decay length takes the form
\begin{eqnarray}\label{phizero}
&& E'_0 = \frac{\lambda+2\sqrt{\alpha} V k_F \sin\delta}{1+\alpha},
\\
&& \kappa =\frac{V}{2t \sqrt{\alpha}} \sin\delta.
\end{eqnarray}
As before we consider the rest of Hamiltonian as perturbation $\delta H = H_1+H_2$. The effective Hamiltonian is then $
H'_{\sigma \sigma'} = \langle\psi_{\sigma }|\delta H|\psi_{\sigma'}\rangle
$
However, the quadratic part ($H_2$) of $\delta H$  no longer cancels exactly and gives the diagonal part of the effective Hamiltonian
\begin{equation}
 \langle\psi_{\sigma}| H_2|\psi_{\sigma '}\rangle = \delta_{\sigma
 \sigma '} 
H_{\uparrow\uparrow}
\end{equation}
where
\begin{equation}
H_{\uparrow\uparrow} =  \langle\psi_{\uparrow}| H_2|\psi_{\uparrow}\rangle =\frac{\alpha k_z^2 }{N^2} \cdot \frac{1- \exp(-2\frac{\kappa \delta}{k_z})}{4\kappa (k_z^2+\kappa^2)} t k_\perp^2,
\end{equation}
where 
\begin{equation}
\mathcal{N}^2 = \frac{k_z^2}{k_z^2+\kappa^2}\frac{\alpha + \exp(-2\frac{\kappa \delta}{k_z})}{4\kappa}
\end{equation}
leading to 
\begin{equation}
H_{\uparrow\uparrow} = t \alpha \cdot \frac{1- \exp(-2\frac{\kappa \delta}{k_z})}{\alpha+  \exp(-2\frac{\kappa \delta}{k_z})}k_\perp^2
\end{equation}

While the linear part gives
\begin{equation}
H_{\uparrow\downarrow } = \langle\psi_{\uparrow}| H_1|\psi_-\rangle = 
2i k_- \frac{V}{k_z}\sqrt{\alpha}\frac{k_z \cos\delta +\kappa\sin\delta}{1+ \alpha \exp(-2\frac{\kappa \delta}{k_z})}.
\end{equation}
The effective Hamiltonian is then
\begin{equation}
H_{eff} =H_0+ \pmat{H_{\up\up}& H_{\up\dw}\cr H_{\up\dw}^*&H_{\up\up}} 
\end{equation}
Combining all the pieces together we can diagonalise the effective Hamiltonian to get the energy: $E_{eff} =E_0' + H_{\up\up} \pm |H_{\up\dw}|$
\begin{equation}
E_{eff} =E_0' \pm 2\frac{V}{k_z}\sqrt{\alpha}\frac{k_z \cos\delta +\kappa\sin\delta}{1+ \alpha \exp(-2\frac{\kappa \delta}{k_z})} k_\perp + \alpha t \cdot \frac{1- \exp(-2\frac{\kappa \delta}{k_z})}{\alpha+  \exp(-2\frac{\kappa \delta}{k_z})}k_\perp^2.
\end{equation}

\bibliographystyle{apsrev}

\end{document}